\documentclass[runningheads]{llncs}
\usepackage{graphicx}
\usepackage{caption}
\usepackage{multirow}
\usepackage{amsmath,amssymb,bm}
\usepackage{bbm}
\usepackage{booktabs}
\usepackage{bbding}
\usepackage{colortbl}
\usepackage[dvipsnames, svgnames, x11names]{xcolor}
\definecolor{citecolor}{HTML}{0071BC}
\usepackage[pagebackref,breaklinks,colorlinks,citecolor=citecolor]{hyperref}
\usepackage{subcaption}

\begin{document}
\title{Towards Generalizable Medical Image Segmentation with Pixel-wise Uncertainty Estimation}
\titlerunning{DG}

%
%
\author{Shuai Wang\inst{1}, Zipei Yan\inst{2}, Daoan Zhang\inst{3}, Zhongsen Li\inst{1} \\
         Sirui Wu\inst{1}, Wenxuan Chen\inst{1}, Rui Li\inst{1}}

\institute{Tsinghua University         
          \and The Hong Kong Polytechnic University
          \and Southern University of Science and Technology
          }
\authorrunning{Wang et al.}
%
%
\maketitle              
\begin{abstract}
Deep neural networks (DNNs) achieve promising performance in visual recognition under the independent and identically distributed (IID) hypothesis. 
In contrast, the IID hypothesis is not universally guaranteed in numerous real-world applications, especially in medical image analysis. 
Medical image segmentation is typically formulated as a pixel-wise classification task in which each pixel is classified into a category.
However, this formulation ignores the hard-to-classified pixels, e.g., some pixels near the boundary area, as they usually confuse DNNs.
In this paper, we first explore that hard-to-classified pixels are associated with high uncertainty.
Based on this, we propose a novel framework that utilizes uncertainty estimation to highlight hard-to-classified pixels for DNNs, thereby improving its generalization.
We evaluate our method on two popular benchmarks: prostate and fundus datasets. 
The results of the experiment demonstrate that our method outperforms state-of-the-art methods.
\end{abstract}

\section{Introduction}
Deep neural networks (DNNs) are state-of-the-art methods in visual recognition, heavily relying on the hypothesis that training and test data are assumed to be sampled from the same distribution, i.e., independent and identically distributed (IID) hypothesis. However, this hypothesis is not universally guaranteed in numerous real-world applications, especially in medical image analysis. Specifically, distribution shift occurs naturally in medical image analysis because medical images from different data sources have distinct imaging modalities and unique acquisition parameters. Therefore, the generalization of DNNs becomes a fundamental problem in medical image analysis, and sometimes pre-trained DNNs unexpectedly predict poorly on out-of-distribution (OOD) samples. To address this problem, domain generalization that aims to generalize to unseen target domains has been proposed.

\begin{figure}[t]
    \centering
    \begin{subfigure}[b]{0.23\textwidth}
        \centering
        \includegraphics[width=\textwidth]{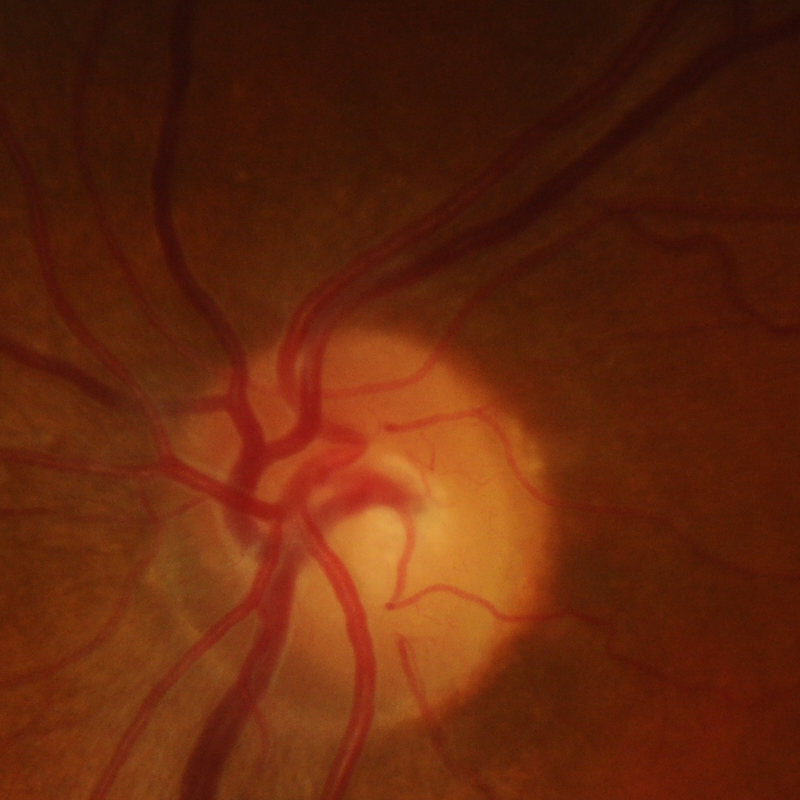}
        \caption{Sample}
    \end{subfigure}
    \begin{subfigure}[b]{0.23\textwidth}
        \centering
        \includegraphics[width=\textwidth]{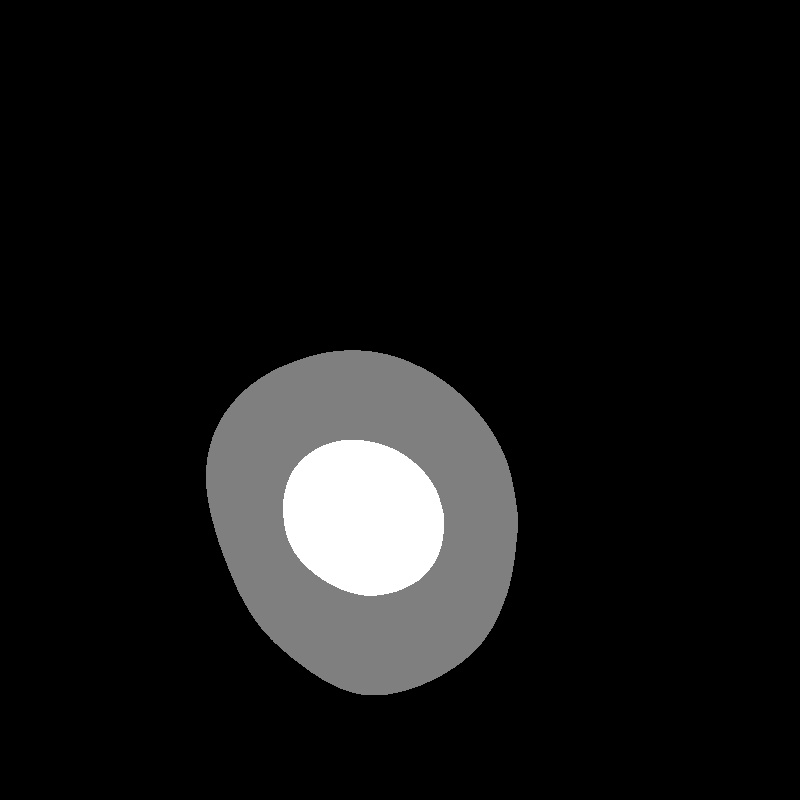}
        \caption{GT}
    \end{subfigure}
    \begin{subfigure}[b]{0.23\textwidth}
        \centering
        \includegraphics[width=\textwidth]{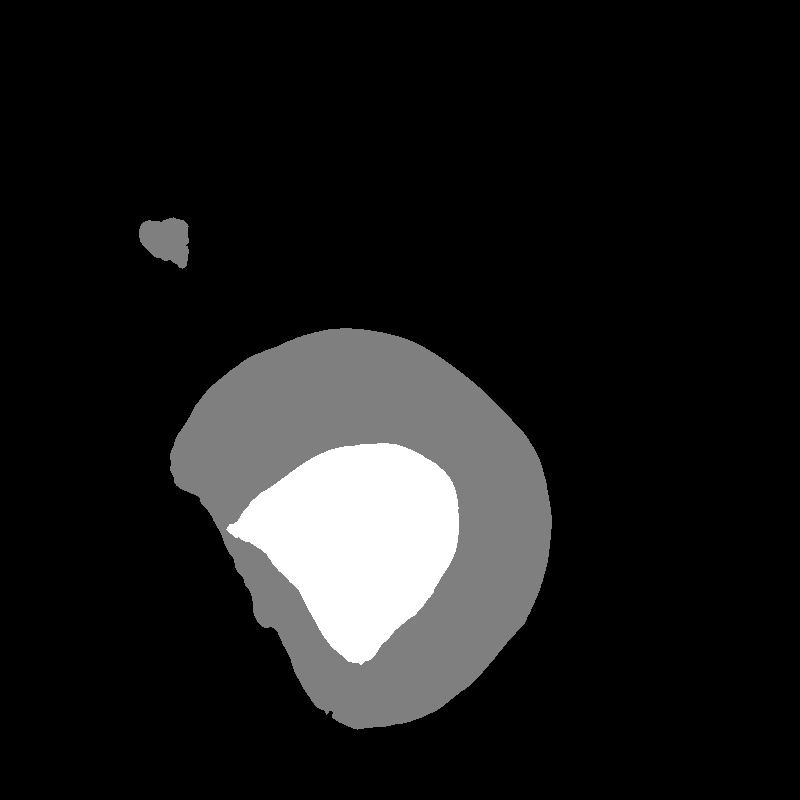}
        \caption{Prediction}
    \end{subfigure}
    \begin{subfigure}[b]{0.23\textwidth}
        \centering
        \includegraphics[width=\textwidth]{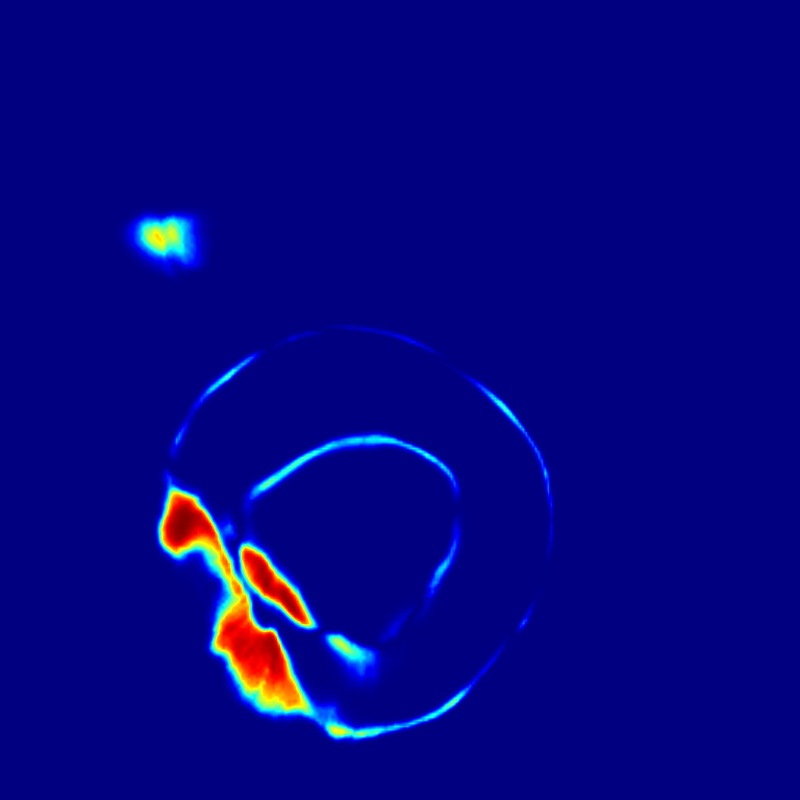}
        \caption{Uncertainty}
    \end{subfigure}
    \begin{subfigure}[b]{0.045\textwidth}
        \centering
        \includegraphics[trim={0.0 13.5 0.0 8.0},clip,width=\textwidth]{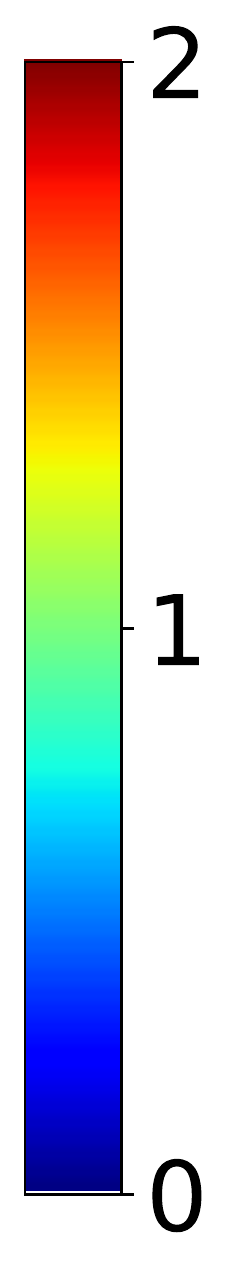}
        \caption*{}
    \end{subfigure}
    \caption{\textbf{The illustration of motivation}. From left to right: (a) an OOD fundus sample, (b) ground truth, (c) prediction from a DNN on (a), (d) uncertainty map of (c).}
  \label{fig:teaser}
\end{figure}

To tackle the domain generalization problem in medical image segmentation, a variety of methods have been explored, including data augmentation \cite{Zhang2020}, self-supervised learning \cite{jigsaw,zhou2022generalizable}, meta-learning \cite{feddg,saml}, and representation learning \cite{dofe}. Specifically, almost all of them follow the  formulation that the segmentation task is performed as a pixel-wise classification task where each pixel is classified into a category. However, this formulation ignores the problem that DNNs are usually confused with \textit{hard-to-classified} pixels, e.g., pixels located in the boundary area. As illustrated in Fig.~\ref{fig:teaser}, we visualize a DNN's prediction on an OOD sample. Specifically, most pixels are predicted \textit{correctly} and \textit{consistently}, except for some \textit{hard-to-classified} pixels. By estimating its uncertainty, we observe that \textit{hard-to-classified} pixels remain highly uncertain. Therefore, this observation motivates us to design a method that could highlight \textit{hard-to-classified} pixels for DNNs thereby improving its performance.

In this paper, we propose a novel framework to tackle the domain generalization problem for medical image segmentation by estimating pixel-wise uncertainty. To be specific, we estimate the pixel-wise uncertainty via Monte Carlo Dropout~\cite{monte_carlo_dropout,gal_uncertainty}, then propose uncertainty-weighted loss function to explicitly highlight the \textit{hard-to-classified} pixels, thereby improving DNN's generalization. To evaluate our method, we conduct extensive experiments on two representative benchmarks, i.e., prostate and fundus datasets. The experimental results demonstrate that our method not only outperforms the baseline by a significant margin but also surpasses recent state-of-the-art methods.

\section{Method}
Let $\{(x_i,y_i)\}\in \mathcal{D}_k$ denotes image-label pairs sampled from domain $\mathcal{D}_k$. The overall objective is to train a DNN $f_{\theta}(\cdot)$ on multiple source domains $\mathcal{D}=\{ \mathcal{D}_1,\mathcal{D}_2,\cdots,\mathcal{D}_K\}$ such that $f_{\theta}(\cdot)$ could generalize to unseen domain $\mathcal{D}_{K+1}$.

\begin{figure}[t]
  \includegraphics[width=\textwidth]{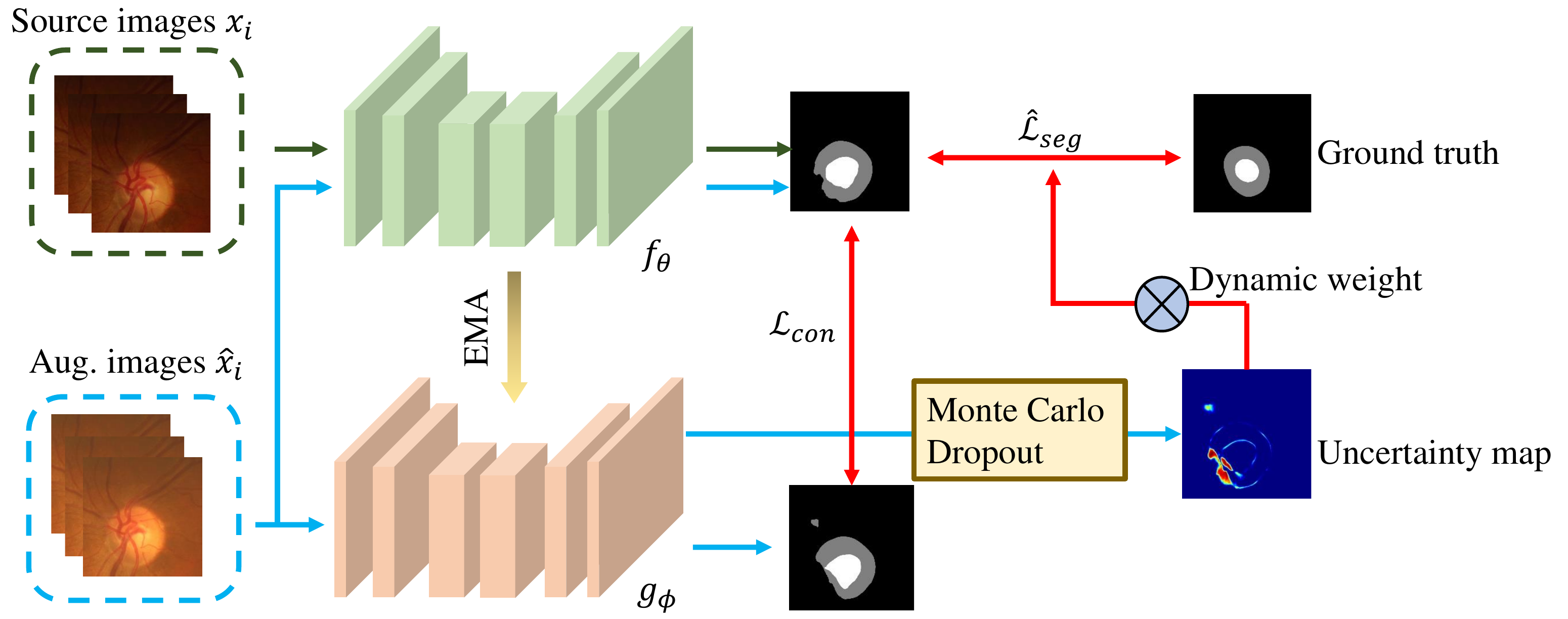}
  \caption{The overview of our method.}
  \label{fig:framework}
\end{figure}

The overview of the proposed method is illustrated in Fig.~\ref{fig:framework}. In the following subsections, we introduce the details of our method.

\subsection{Preliminaries}
Given an image $x_i\in \mathbb{R}^{H\times W\times C}$ where $H$/$W$ stands height/width and $C$ denotes the number of channels, and its ground-truth label $y_i \in \mathbb{R}^{H\times W}$. The hybrid segmentation loss~\cite{nnunet} combines the Dice loss and binary cross-entropy loss, which is defined as follows
\begin{equation}
    \mathcal{L}_{seg}(p_i, y_i) = \mathcal{L}_{dice}(p_i, y_i) + \mathcal{L}_{ce}(p_i,y_i),
    \label{eq:seg_loss}
\end{equation}
where $\mathcal{L}_{dice}$ and $\mathcal{L}_{ce}$ are defined as follows:
\begin{equation}
    \mathcal{L}_{dice}(p_i, y_i) = 1 - \frac{2\sum_{n=1}^{HW} p_i^n y_i^n}{\sum_{n=1}^{HW} (p_i^n)^2+(y_i^n)^2},
    \label{eq:loss_dice}
\end{equation}

\begin{equation}
  \mathcal{L}_{ce}(p_i,y_i) = -\frac{1}{HW}\sum_{n=1}^{HW} (y_i^n\log p_i^n + (1-y_i^n)\log(1-p_i^n)),
  \label{eq:loss_ce}
\end{equation}
where $p_i=f_{\theta}(x_i)$ denotes the predicted probabilistic distribution map of $x_i$. Besides, $p_i^n$ and $y_i^n$ denote the $n$-th element from $p_i$ and $y_i$, separately. 

\noindent
\textbf{Remark}.  The above objective function optimizes pixel-wise classification, where each pixel is treated equally, thereby ignoring paying more attention to \textit{hard-to-classified} pixels. As illustrated early, we point out that these \textit{hard-to-classified} pixels deserve more attention as they usually confuse the DNNs.
\subsection{Fourier-based Data augmentation}
DNNs are sensitive to the variation of image styles~\cite{GeirhosRMBWB19}. To make DNNs more robust to variation of image styles, we adopt the Fourier-based data augmentation~\cite{feddg,fourier_dg,fourier_da} to augment $x_i$ to $\hat{x}_i$ that shares the same semantic but different style property. The core insight behind this is that the phase component of the Fourier spectrum preserves the high-level semantics of the original signal, while the amplitude component contains low-level statistics~\cite{import_phase,fourier_dg,fourier_da}. For an image $x_i$, its frequency space signal $\mathcal{F}(x_i)$ can be obtained with Fast Fourier Transform (FFT), which is defined as follows
\begin{equation}
    \mathcal{F}(x_i)(u,v,c) = \sum_{h=1}^{H}\sum_{w=1}^{W} x_i(h,w,c) e^ {-j2\pi(\frac{h}{H}u+\frac{w}{W}v)}=\mathcal{A}(x_i)e^{j\mathcal{P}(x_i)},
\end{equation}
where $\mathcal{A}(x_i)$ and $\mathcal{P}(x_i)$ denote amplitude and phase spectrum of $x_i$, respectively. 

To augment an image $x_i$, we disturb $\mathcal{A}(x_i)$ to change its style but preserve  $\mathcal{P}(x_i)$ to keep its semantic information. Specifically, we randomly sample another image $x^{\prime}$ to obtain its $\mathcal{A}(x^{\prime})$, then mix it with $\mathcal{A}(x_i)$ as follows
\begin{equation}
    \mathcal{A}(\hat{x}_i) = (\lambda \mathcal{A}(x^{\prime}) + (1-\lambda)\mathcal{A}(x_i)) \odot \mathcal{M} + \mathcal{A}(x_i) \odot (1-\mathcal{M}),
\end{equation}
where $\lambda \sim \text{Beta}(\alpha,\alpha)$, $\mathcal{M}=\mathbbm{1}_{(h,w)\in [-\alpha H:\alpha H,-\alpha W:\alpha W]}$ denotes a binary mask, $\odot$ denotes Hadamard product. In practice, we set $\alpha=0.1$ during the training process.

Thereafter, we recombine $\mathcal{A}(\hat{x_i})$ and original phase $\mathcal{P}(x_i)$ to generate an augmented image $\hat{x}_i$ with inverse FFT $\mathcal{F}^{-1}$ as
\begin{equation}
  \hat{x}_i = \mathcal{F}^{-1}(\mathcal{A}(\hat{x}_i) e^{j \mathcal{P}(x_i)}).
\end{equation}
Finally, we obtain augmented $\hat{x}_i$ which shares the same semantic information but in different styles with $x_i$. 

\subsection{Momentum-updated Network}
To reduce the optimized parameter fluctuation noise and provide long-term memory~\cite{mean-teacher}, we introduce a momentum-updated network $g_{\phi}(\cdot)$ by Exponential Moving Average (EMA) ~\cite{mean-teacher} from existing $f_{\theta}(\cdot)$, where its parameters are updated as follows
\begin{equation}
    \phi = m\phi + (1-m)\theta,
    \label{eq:momentum_update}
\end{equation}
where $m$ is the momentum parameter that controls the updating rate.

Besides, we introduce a consistency loss to force both $g_{\phi}(\cdot)$ and $f_{\theta}(\cdot)$ to predict consistently on a given $x_i$ and its augmented $\hat{x}_i$, which is defined as follows
\begin{equation}
    \mathcal{L}_{con}(f_{\theta}(x_i), g_{\phi}(\hat{x}_i)) = \frac{1}{HW} \sum_{n=1}^{HW} D_{\text{KL}}(f_{\theta}(x_i)^n \| g_{\phi}(\hat{x}_i)^n),
    \label{eq:loss_con}
\end{equation}
where $D_{\text{KL}}(\cdot \| \cdot)$ denotes the Kullback-Leibler divergence.

\subsection{Pixel-wise Uncertainty Estimation}
Our key idea is to highlight \textit{hard-to-classified} pixels for DNNs, such that they could pay more attention to them, thereby improving its generalization.

We first estimate pixel-wise uncertainty with Monte Carlo Dropout~\cite{monte_carlo_dropout,gal_uncertainty}. Specifically, we perform $T$ stochastic forward passes through $g_{\phi}(\cdot)$ on $\hat{x}_i$ with random noise injection and dropout, then calculate its entropy of predictions as the uncertainty map, which is defined as follows
\begin{equation}
    \overline{u_i} = \frac{1}{T}\sum_{k=1}^{T}g_{\phi}(\hat{x}_i + \epsilon_k), u_i = -\sum \overline{u_i} \log \overline{u_i},
\end{equation}
where $\epsilon_k \sim \mathcal{N}(0,\sigma^2)$. We set $\sigma=0.1$ and $T=8$ by default. 

Then, to explicitly highlight \textit{hard-to-classified} pixels, we propose the uncertainty weighted binary cross-entropy loss incorporating the above uncertainty as a dynamic weight, which is formulated as follows
\begin{equation}
  \mathcal{L}_{uce}(p_i, y_i, u_i) = -\frac{1}{HW}\sum_{n=1}^{HW} u_i^n (y_i^n\log p_i^n + (1-y_i^n)\log(1-p_i^n)).
  \label{eq:loss_wce}
\end{equation}
Thereafter, the hybrid segmentation loss can be redefined as follows
\begin{equation}
    \mathcal{\hat{L}}_{seg}(p_i, y_i, u_i) = \mathcal{L}_{dice}(p_i, y_i) + \mathcal{L}_{uce}(p_i,y_i, u_i).
    \label{eq:new_seg_loss}
\end{equation}

\subsection{Overall Objective Function}
The overall objective function of our method is formulated as follows
\begin{equation}
    \mathcal{L}(\theta, \phi) = \mathcal{\hat{L}}_{seg}(f_{\theta}(x_i), y_i,u_i) + \mathcal{\hat{L}}_{seg}(f_{\theta}(\hat{x}_i), y_i,u_i) + \beta\mathcal{L}_{con}(f_{\theta}(x_i), g_{\phi}(\hat{x}_i)),
    \label{eq:loss_overall}
\end{equation}
where $\beta$ is the hyper-parameter to control the importance of consistency loss.

\section{Experiments}
In this section, we conduct experiments on two representative datasets to evaluate the effectiveness of our method. Besides we conduct ablation studies to examine the effectiveness of different components in our method.

\subsection{Datasets}
\textbf{Prostate} dataset\footnote{\href{https://liuquande.github.io/SAML/}{https://liuquande.github.io/SAML/}} contains 116 T2-weighted MRI volumes from six different domains \cite{saml,lemaitre2015computer,litjens2014evaluation}. Each domain contains 30/30/19/13/12/12 volumes, respectively. We resize all volumes to $384\times 384$ resolution and we use 2d slices for training following common practice~\cite{saml,zhou2022generalizable}.

\textbf{Fundus} dataset\footnote{\href{https://drive.google.com/file/d/1p33nsWQaiZMAgsruDoJLyatoq5XAH-TH/view}{https://drive.google.com/file/d/1p33nsWQaiZMAgsruDoJLyatoq5XAH-TH/view}}~\cite{dofe} includes retina fundus images from four different clinical centers. This dataset is composed of 3 public datasets including Drishti-GS dataset~\cite{sivaswamy2015comprehensive}, RIM-ONE-r3 dataset~\cite{fumero2011rim} and REFUGE dataset~\cite{Orlando2020}. Each domain contains 101/159/400/400 2D images, respectively. We follow data split and pre-processing in \cite{dofe,zhou2022generalizable}. Besides, we crop images with an $800 \times 800$ bounding box and resize all images to $256 \times 256$ resolution. In addition, we normalize all images to $[-1,1]$.

\subsection{Implementation Details}
We employ the 2D UNet~\cite{unet} as the segmentation network following~\cite{zhou2022generalizable}. Notably, our method is also model architecture agnostic, which can be further utilized for other models. As for training details, we optimize the network with Adam optimizer~\cite{adam}. Specifically, we set the learning rate to $1e^{-4}$ with a batch size of 8 and train 500/200 epochs for the Prostate dataset and Fundus dataset, respectively. In our experiments, we follow the common practice of domain generalization and employ the \textit{leave-one-domain-out} strategy. To be more specific, we train the model on images from $K$ source domains and evaluate it on the $K+1$ domain.

For all experiments, the momentum $m$ (\textit{cf}. Eq. \ref{eq:momentum_update}) is set to 0.99 and $\beta$ (\textit{cf}. Eq.~\ref{eq:loss_overall}) is set to 200. We investigate their effect in Table~\ref{table:setting_beta} and \ref{table:setting_m}. For $\beta$, we also use sigmoid ramp-up~\cite{mean-teacher} for $\beta$ with $\frac{1}{10}$ of the number of epochs during training. We use the prediction from  $f_{\theta}(\cdot)$ for evaluation. We employ two commonly-used metrics in medical image segmentation: Dice Score (DSC) and Average Surface Distance (ASD). Specifically, DSC measures the overlap between prediction and ground truth, while ASD measures the performance at the object boundary. Note that higher DSC and lower ASD indicate better performance.

\begin{table}[t]
    \centering
    \caption{\textbf{Results on Prostate dataset.} The best results are \textbf{bold-faced}, and the second-best results are \underline{underlined}. The results of first block is from~\cite{zhou2022generalizable}.}
    \resizebox{\textwidth}{!}{
      \begin{tabular}{l|ccccccc|ccccccc}
      \toprule
      Metric & \multicolumn{7}{c|}{DSC (\textuparrow)}                              & \multicolumn{7}{c}{ASD (\textdownarrow)} \\
      \midrule
      Domain & A     & B     & C     & D     & E     & F     & Avg   & A     & B     & C     & D     & E     & F     & Avg \\
      \midrule
      JiGen~\cite{jigsaw} & 85.45 & 89.26 & 85.92 & 87.45 & 86.18 & 83.08 & 86.22 & 1.11  & 1.81  & 2.61  & 1.66  & \textbf{1.71} & 2.43  & 1.89 \\
      BigAug~\cite{Zhang2020} & 85.73 & 89.34 & 84.49 & 88.02 & 81.95 & 87.63 & 86.19 & 1.13  & 1.78  & 4.01  & 1.25  & 1.92  & 1.89  & 2.00 \\
      SAML~\cite{saml}  & 86.35 & 90.18 & 85.03 & 88.20  & 86.97 & 87.69 & 87.40  & 1.09  & 1.54  & 2.52  & 1.41  & 2.01  & 1.77  & 1.72 \\
      FedDG~\cite{feddg} & 86.43 & 89.59 & 85.30  & 88.95 & 85.93 & 87.39 & 87.27 & 1.30   & 1.67  & 2.36  & 1.37  & 2.19  & 1.94  & 1.81 \\
      DoFE~\cite{dofe}  & 89.64 & 87.56 & 85.08 & 89.06 & 86.15 & 87.03 & 87.42 & \textbf{0.92} & 1.49  & 2.74  & 1.46  & 1.89  & 1.53  & 1.68 \\
      DSIR~\cite{zhou2022generalizable}  & 87.56 & \underline{90.20}  & \underline{86.92} & 88.72 & \underline{87.17} & 87.93 & 88.08 & \underline{1.04}  & \textbf{0.81} & \underline{2.23}  & 1.16  & \underline{1.81}  & 1.15  & \underline{1.37} \\
      \midrule
      ERM~\cite{vapnik_stat_learn_theory} & 89.18 & 85.92 & 81.26 & 87.44 & 74.95 & 86.37 & 84.19 & 1.70   & 1.56  & 3.68  & 1.72  & 4.83  & 1.91  & 2.57 \\
      CutMix~\cite{cutmix} & 90.17 & 85.23 & 82.46 & 89.85 & 72.80 & 90.52 & 85.30 & 1.26 &1.53 & 2.70 & 1.03 & 6.36 & 0.86 & 2.29 \\
      Mixup~\cite{mixup} & \underline{91.31} & 88.25 & 85.91 & \underline{90.16} & 84.13 & \underline{91.16} & \underline{88.49} & \underline{1.04} & 1.17 & 2.25 & \underline{1.02} & 2.25 & \underline{0.77} & 1.42 \\
      \rowcolor{gray!25}
      Ours & \textbf{91.34} & \textbf{91.22} & \textbf{88.38} & \textbf{90.33} & \textbf{89.38} & \textbf{91.50} & \textbf{90.36} & 1.28  & \underline{0.91}  & \textbf{1.61} & \textbf{0.98} & 1.89  & \textbf{0.68} & \textbf{1.23} \\
      \bottomrule
      \end{tabular}}
    \label{tab:results_prostate}
  \end{table}
  
\subsection{Baseline Methods}
We compare our method with nine baseline methods, which include \textbf{ERM}\cite{vapnik_stat_learn_theory}: the empirical risk minimization baseline; \textbf{Jigen}~\cite{jigsaw}: a self-supervise learning method by solving jigsaw puzzles; \textbf{BigAug}~\cite{Zhang2020}: a data augmentation method designed for medical image segmentation task; \textbf{FedDG}~\cite{feddg} and \textbf{SAML}~\cite{saml}: two meta-learning based methods; \textbf{DoFE}~\cite{dofe}: a domain-invariant representation learning method; \textbf{DSIR}~\cite{zhou2022generalizable}: a recent state-of-the-art method combing amplitude mixup and self-supervised learning; \textbf{CutMix}~\cite{cutmix} and \textbf{Mixup}~\cite{mixup}: two data augmentation methods to regularize deep neural networks.
\subsection{Experimental Results}

\textbf{Results on Prostate dataset} are reported in Table~\ref{tab:results_prostate}. In general, our method achieves the best performance according to the Avg of DSC and ASD. Compare to the ERM~\cite{vapnik_stat_learn_theory} baseline, our method achieves consistent and significant improvement. Furthermore, our method outperforms the recent state-of-the-art method DSIR~\cite{zhou2022generalizable} in terms of DSC (2.28\%) and ASD (0.14) over six domains, respectively.

\textbf{Results on Fundus dataset} are reported in Table~\ref{tab:results_fundus}. We observe that our method still outperforms the ERM~\cite{vapnik_stat_learn_theory} baseline consistently. In addition, compared to DSIR~\cite{zhou2022generalizable}, our method achieves the best Avg DSC and second-best Avg ASD, which further illustrates the effectiveness of our method.

\begin{figure}[t]
    \begin{center}
        \begin{subfigure}[t]{0.15\linewidth}
            \centering
            \includegraphics[width=0.99\linewidth]{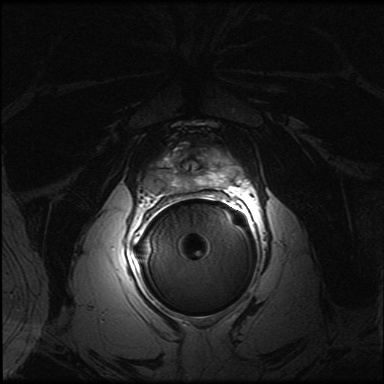}
            \caption*{}            
        \end{subfigure}
        \hfill
        \begin{subfigure}[t]{0.15\linewidth}
            \centering
            \includegraphics[width=0.99\linewidth]{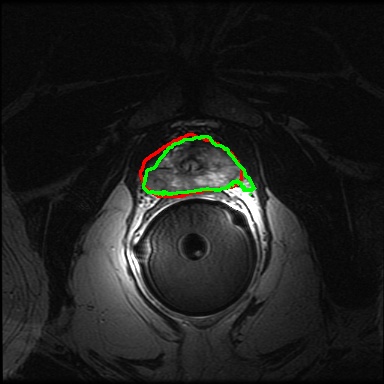}
            \caption*{}            
        \end{subfigure}
        \hfill
        \begin{subfigure}[t]{0.15\linewidth}
            \centering
            \includegraphics[width=0.99\linewidth]{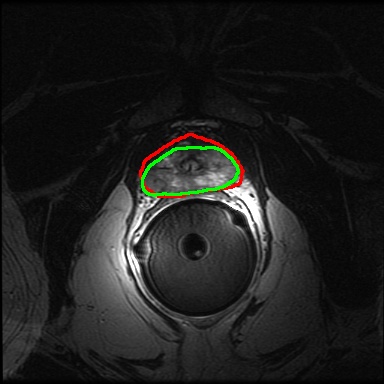}
            \caption*{}            
        \end{subfigure}
        \hfill
        \begin{subfigure}[t]{0.15\linewidth}
            \centering
            \includegraphics[width=0.99\linewidth]{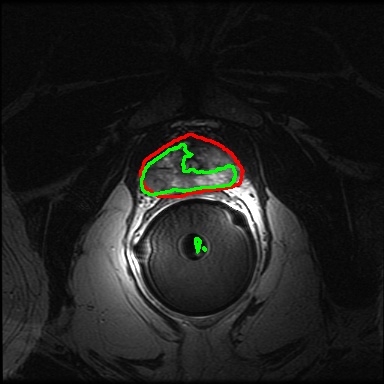}
            \caption*{}            
        \end{subfigure}
        \hfill
        \begin{subfigure}[t]{0.15\linewidth}
            \centering
            \includegraphics[width=0.99\linewidth]{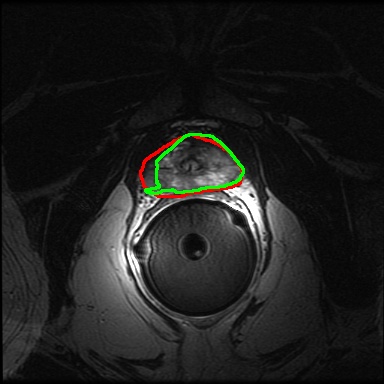}
            \caption*{}            
        \end{subfigure}
        \hfill
        \begin{subfigure}[t]{0.15\linewidth}
            \centering
            \includegraphics[width=0.99\linewidth]{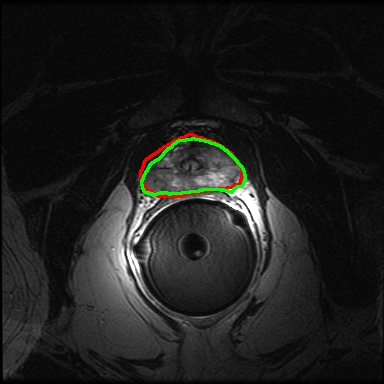}
            \caption*{}            
        \end{subfigure}
        
        \begin{subfigure}[t]{0.15\linewidth}
            \centering
            \includegraphics[width=0.99\linewidth]{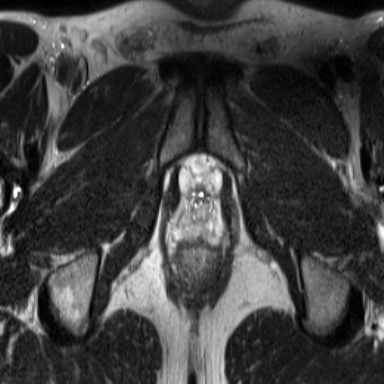}
            \caption*{}            
        \end{subfigure}
        \hfill
        \begin{subfigure}[t]{0.15\linewidth}
            \centering
            \includegraphics[width=0.99\linewidth]{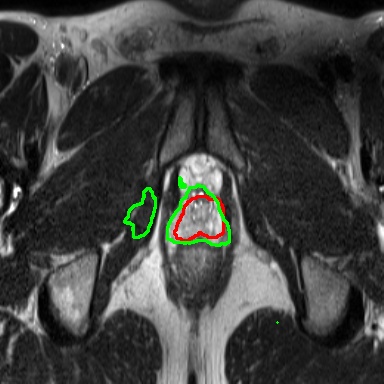}
            \caption*{}            
        \end{subfigure}
        \hfill
        \begin{subfigure}[t]{0.15\linewidth}
            \centering
            \includegraphics[width=0.99\linewidth]{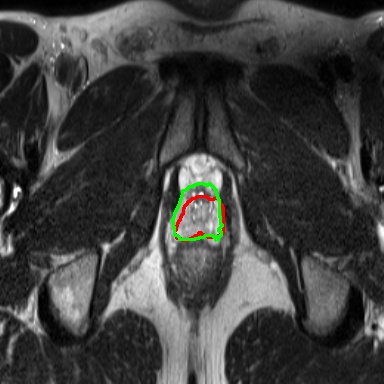}
            \caption*{}            
        \end{subfigure}
        \hfill
        \begin{subfigure}[t]{0.15\linewidth}
            \centering
            \includegraphics[width=0.99\linewidth]{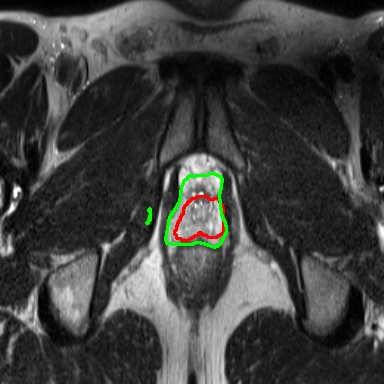}
            \caption*{}            
        \end{subfigure}
        \hfill
        \begin{subfigure}[t]{0.15\linewidth}
            \centering
            \includegraphics[width=0.99\linewidth]{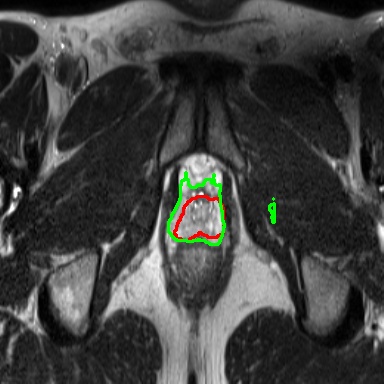}
            \caption*{}            
        \end{subfigure}
        \hfill
        \begin{subfigure}[t]{0.15\linewidth}
            \centering
            \includegraphics[width=0.99\linewidth]{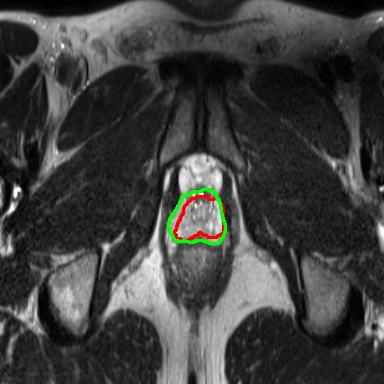}
            \caption*{}            
        \end{subfigure}

    \begin{subfigure}[t]{0.15\linewidth}
            \centering
            \includegraphics[width=0.99\linewidth]{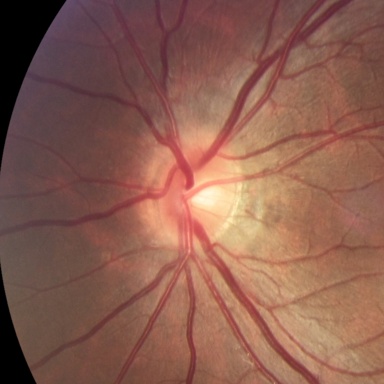}
            \caption*{}            
        \end{subfigure}
        \hfill
        \begin{subfigure}[t]{0.15\linewidth}
            \centering
            \includegraphics[width=0.99\linewidth]{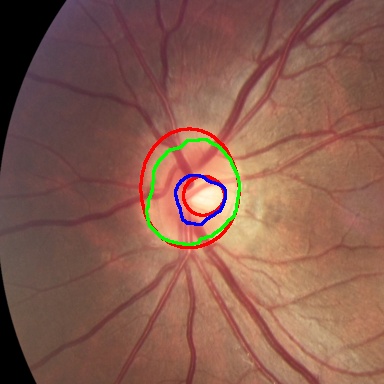}
            \caption*{}            
        \end{subfigure}
        \hfill
        \begin{subfigure}[t]{0.15\linewidth}
            \centering
            \includegraphics[width=0.99\linewidth]{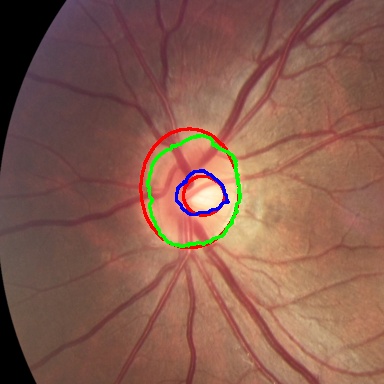}
            \caption*{}            
        \end{subfigure}
        \hfill
        \begin{subfigure}[t]{0.15\linewidth}
            \centering
            \includegraphics[width=0.99\linewidth]{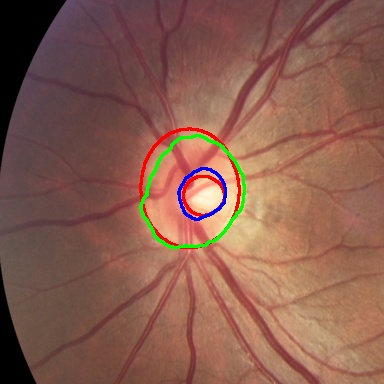}
            \caption*{}            
        \end{subfigure}
        \hfill
        \begin{subfigure}[t]{0.15\linewidth}
            \centering
            \includegraphics[width=0.99\linewidth]{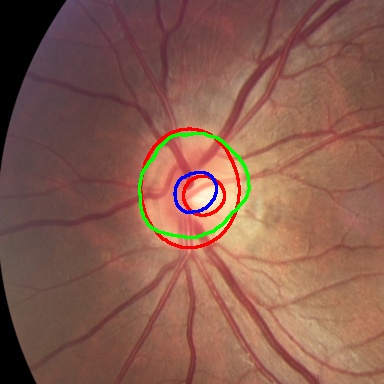}
            \caption*{}            
        \end{subfigure}
        \hfill
        \begin{subfigure}[t]{0.15\linewidth}
            \centering
            \includegraphics[width=0.99\linewidth]{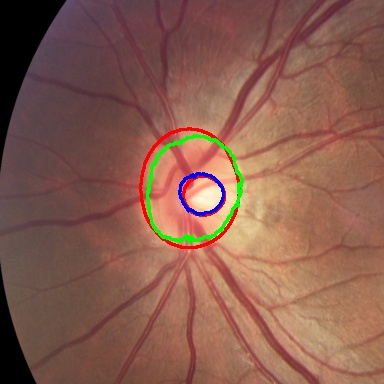}
            \caption*{}            
        \end{subfigure}

    \begin{subfigure}[t]{0.15\linewidth}
            \centering
            \includegraphics[width=0.99\linewidth]{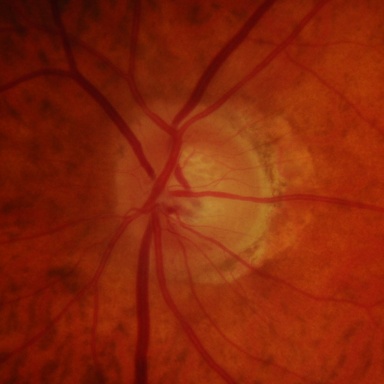}
            \caption*{Image}            
        \end{subfigure}
        \hfill
        \begin{subfigure}[t]{0.15\linewidth}
            \centering
            \includegraphics[width=0.99\linewidth]{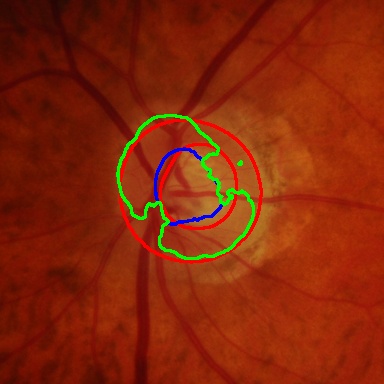}
            \caption*{ERM}            
        \end{subfigure}
        \hfill
        \begin{subfigure}[t]{0.15\linewidth}
            \centering
            \includegraphics[width=0.99\linewidth]{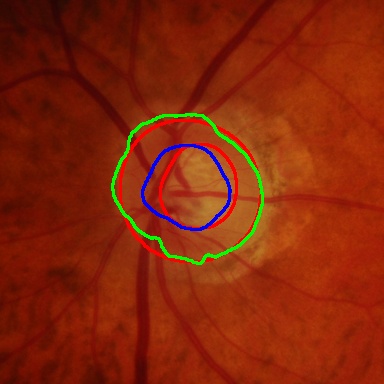}
            \caption*{Mixup}            
        \end{subfigure}
        \hfill
        \begin{subfigure}[t]{0.15\linewidth}
            \centering
            \includegraphics[width=0.99\linewidth]{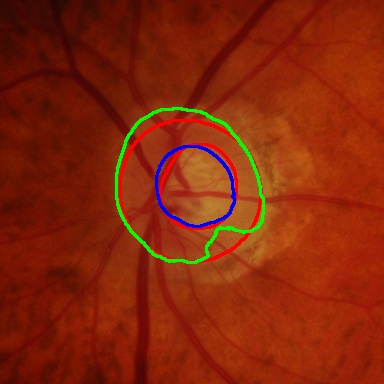}
            \caption*{CutMix}            
        \end{subfigure}
        \hfill
        \begin{subfigure}[t]{0.15\linewidth}
            \centering
            \includegraphics[width=0.99\linewidth]{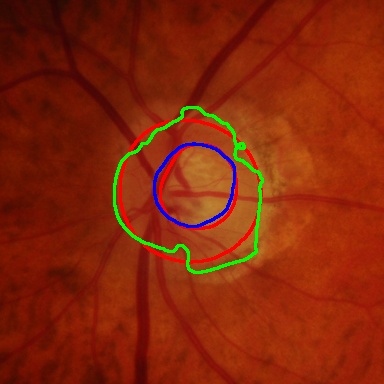}
            \caption*{DSIR}            
        \end{subfigure}
        \hfill
        \begin{subfigure}[t]{0.15\linewidth}
            \centering
            \includegraphics[width=0.99\linewidth]{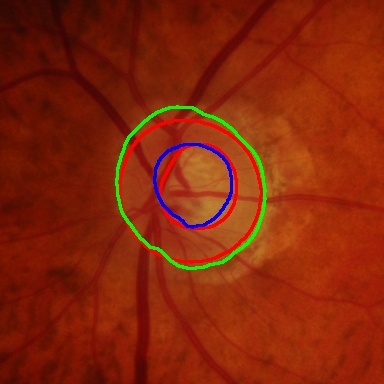}
            \caption*{Ours}            
        \end{subfigure}
    \end{center}
    \vspace{-15pt}
    \caption{Visualization of predicted boundaries from different methods on \textbf{Prostate} MRI and \textbf{Fundus} datasets. The first two rows present results on the \textbf{Prostate} dataset where red contours indicate the boundary of ground truth while blue and green contours represent prediction of optic cup and optic disk, respectively. And last two rows are for \textbf{Fundus} dataset where green and red contours indicate the boundary of prediction and ground truth, respectively.}
  \label{fig:vis_results}
\end{figure}

\textbf{Visualization of predictions.} We also present qualitative results in Fig. \ref{fig:vis_results}. In general, we observe that our method could produce better segmentation boundaries while other methods may generate misleading ones. 

\begin{table}[t]
    \centering
    \caption{\textbf{Results on Fundus dataset.} The best results are \textbf{bold-faced}, and the second-best results are \underline{underlined}.The results of first block is from~\cite{zhou2022generalizable}.}
    \begin{tabular}{l|ccccc|ccccc}
        \toprule
        Metric & \multicolumn{5}{c|}{DSC (\textuparrow)}              & \multicolumn{5}{c}{ASD (\textdownarrow)} \\
        \midrule
        Domain & A     & B     & C     & D     & Avg   & A     & B     & C     & D     & Avg \\
        \midrule
        JiGen~\cite{jigsaw} & 88.74 & 82.15 & \underline{90.98} & 86.11 & 86.99 & 14.00    & 18.41 & \textbf{8.07}  & 13.99 & 13.62 \\
        BigAug~\cite{Zhang2020} & 85.50  & 81.55 & 88.01 & 86.92 & 85.49 & 17.57 & 17.80 & 10.91 & 10.47 & 14.18 \\
        SAML~\cite{saml}  & 89.38 & 82.63 & 89.35 & 87.43 & 87.19 & 13.05 & 17.68 & 9.37  & 12.46 & 13.14 \\
        FedDG~\cite{feddg} & 88.67 & 83.29 & 89.40 & 87.64 & 87.25 & 13.13 & 16.40  & 9.19 & 8.87  & 11.90 \\
        DoFE~\cite{dofe}  & 89.57 & \textbf{85.16} & 89.11 & \textbf{90.16} & 88.50 & 11.63 & 15.25 & 10.44 & \underline{7.72} & 11.26 \\
        DSIR~\cite{zhou2022generalizable}  & 90.62 & 84.13 & \textbf{91.06} & 89.97 & \underline{88.94} & \underline{11.59} & \textbf{13.94} & \textbf{8.07} & \textbf{7.68} & \textbf{10.32} \\
          \midrule
        ERM~\cite{vapnik_stat_learn_theory}   & 87.81 & 79.70  & 89.21 & 82.59 & 84.83 & 11.70  & 18.60  & 10.66 & 11.53 & 13.12 \\
        CutMix~\cite{cutmix} & \underline{91.09} & 82.52 &89.55 & 89.71 & 88.22 & 11.07 & 17.91 & 10.11 & 7.64 & 11.68 \\
        Mixup~\cite{mixup} & 90.32 & 81.56 & 89.00 & 86.93 & 86.95 & 12.35 & 19.25 & 10.29 & 11.27 & 13.29 \\
        \rowcolor{gray!25} Ours  & \textbf{91.94} & \underline{84.98} & 90.14 & \underline{89.99} & \textbf{89.26} & \textbf{9.84} & \underline{14.22} & \underline{9.24}  & 8.80   & \underline{10.53} \\
        \bottomrule
    \end{tabular}
    \label{tab:results_fundus}
  \end{table}

\subsection{Ablation Study}
\textbf{Effectiveness of different components.} We conduct an ablation study to evaluate the effectiveness of different components in our method. As reported in Table~\ref{tab:ablation_study}, we observe that all components can effectively improve performance. Specifically, Fourier-based data augmentation improves baseline by about 3.07\% and 1.71\% on two benchmarks. Besides, consistency regularization (\textit{cf.} Eq. \ref{eq:loss_con}) brings about 1.14\% and 1.76\% gains. Moreover, our proposed uncertainty-weighted binary cross-entropy loss (\textit{cf.} Eq. \ref{eq:loss_wce}) can further improve performance to 90.36\% and 89.26\% on two benchmarks.
\begin{table}[h]
    \centering
    \setlength{\tabcolsep}{4pt}
    \caption{\textbf{Ablation study} on two benchmarks in terms of DSC metric.}
    \begin{tabular}{c|cc|cc|c|cc}
        \toprule
        \# & $\mathcal{L}_{seg}(f_{\theta}(x))$ & $\mathcal{L}_{seg}(f_{\theta}(\hat{x}))$ & $\mathcal{\hat{L}}_{seg}(f_{\theta}(x))$ & $\mathcal{\hat{L}}_{seg}(f_{\theta}(\hat{x}))$ & $\mathcal{L}_{con}$ & Prostate & Fundus \\
        \midrule
        0 & \Checkmark & & & & & 84.19 & 84.83 \\
        1 & \Checkmark & \Checkmark & & & & 87.26 & 86.54 \\
        2 & \Checkmark & \Checkmark & & & \Checkmark & 88.40 & 88.30 \\
        3 & & & \Checkmark & \Checkmark & \Checkmark & \textbf{90.36} & \textbf{89.26} \\
        \bottomrule
    \end{tabular}
    \label{tab:ablation_study}
\end{table}

\textbf{Impact of hyper-parameters.} We further study the impact of two hyper-parameters: $\beta$ (\textit{cf}. Eq. \ref{eq:loss_overall}) and $m$ (\textit{cf}. Eq. \ref{eq:momentum_update}) on \textbf{Prostate} dataset. For hyper-parameter $\beta$, we choose $\beta \in \{1, 10, 100, 200, 400\}$. As reported in Table \ref{table:setting_beta}, we observe that $\beta = 200$ achieves the best performance among these candidates. For hyper-parameter $m$, we choose $m \in \{0.9, 0.99, 0.995, 0.999 \}$. As reported in Table \ref{table:setting_m}, $m=0.99$ is better than others.
\begin{table}[t]
    \begin{minipage}{0.5\textwidth}
    \caption{Impact of different $\beta$ (\textit{cf}. Eq. \ref{eq:loss_overall}) on 
    \textbf{Prostate} dataset.} \label{tab:setting_beta}
        \begin{center}
        \resizebox{\textwidth}{!}{\begin{tabular}{cccccc}
            \toprule
            $\beta$     & 1     & 10    & 100   & 200   & 400 \\
            \midrule
            DSC (\%)  & 88.96 & 89.29 & 89.69 & \textbf{90.36} & 89.74 \\
            \bottomrule
            \end{tabular}
        } 
        
        \end{center}
        \label{table:setting_beta}
    \end{minipage}  
    \hfill
    \begin{minipage}{0.42\textwidth}
      \caption{Impact of different $m$ (\textit{cf}. Eq. \ref{eq:momentum_update}) on \textbf{Prostate} dataset.}
      \begin{center}
      \resizebox{\textwidth}{!}{
        \begin{tabular}{ccccc}
    \toprule
    $m$ & 0.9    & 0.99   & 0.995   & 0.999 \\
    \midrule
    DSC (\%)  & 89.4 & \textbf{90.36} & 90.12 & 89.78 \\
    \bottomrule
    \end{tabular}      
      }    
    \end{center}
    \label{table:setting_m}
    \end{minipage}
\end{table}
\section{Conclusion}
In this paper, we propose a novel method to tackle the domain generalization problem in medical image segmentation. We first explore that \textit{hard-to-classified} pixels are typically associated with high uncertainty. Based on this, we propose to explicitly highlight these \textit{hard-to-classified} pixels with uncertainty-weighted loss. To evaluate the proposed method, we have conducted experiments on two popular benchmarks. And the experimental results demonstrate the effectiveness of our method, which not only significantly outperforms the ERM baseline but also outperforms recent state-of-the-art methods. Moreover, we have further conducted an in-depth ablation study to better understand the effectiveness of different components in our method. 
\bibliographystyle{splncs04.bst}
\bibliography{reference}

\begin{thebibliography}{10}
\providecommand{\url}[1]{\texttt{#1}}
\providecommand{\urlprefix}{URL }
\providecommand{\doi}[1]{https://doi.org/#1}

\bibitem{jigsaw}
Carlucci, F.M., D'Innocente, A., Bucci, S., Caputo, B., Tommasi, T.: Domain
  generalization by solving jigsaw puzzles. In: CVPR (2019)

\bibitem{fumero2011rim}
Fumero, F., Alay{\'o}n, S., Sanchez, J.L., Sigut, J., Gonzalez-Hernandez, M.:
  Rim-one: An open retinal image database for optic nerve evaluation. In: 2011
  24th international symposium on computer-based medical systems (CBMS).
  pp.~1--6. IEEE (2011)

\bibitem{monte_carlo_dropout}
Gal, Y., Ghahramani, Z.: Dropout as a bayesian approximation: Representing
  model uncertainty in deep learning. In: Proceedings of The 33rd International
  Conference on Machine Learning. Proceedings of Machine Learning Research,
  vol.~48, pp. 1050--1059. PMLR, New York, New York, USA (20--22 Jun 2016)

\bibitem{GeirhosRMBWB19}
Geirhos, R., Rubisch, P., Michaelis, C., Bethge, M., Wichmann, F.A., Brendel,
  W.: Imagenet-trained cnns are biased towards texture; increasing shape bias
  improves accuracy and robustness. In: ICLR (2019)

\bibitem{nnunet}
Isensee, F., Jaeger, P.F., Kohl, S.A.A., Petersen, J., Maier-Hein, K.H.:
  {nnU}-net: a self-configuring method for deep learning-based biomedical image
  segmentation. Nature Methods  \textbf{18}(2),  203--211 (Dec 2020)

\bibitem{gal_uncertainty}
Kendall, A., Gal, Y.: What uncertainties do we need in bayesian deep learning
  for computer vision? In: Advances in Neural Information Processing Systems
  (2017)

\bibitem{adam}
Kingma, D.P., Ba, J.: Adam: {A} method for stochastic optimization. In: ICLR
  (2015)

\bibitem{lemaitre2015computer}
Lema{\^\i}tre, G., Mart{\'\i}, R., Freixenet, J., Vilanova, J.C., Walker, P.M.,
  Meriaudeau, F.: Computer-aided detection and diagnosis for prostate cancer
  based on mono and multi-parametric mri: a review. Computers in biology and
  medicine  \textbf{60},  8--31 (2015)

\bibitem{litjens2014evaluation}
Litjens, G., Toth, R., Van De~Ven, W., Hoeks, C., Kerkstra, S., van Ginneken,
  B., Vincent, G., Guillard, G., Birbeck, N., Zhang, J., et~al.: Evaluation of
  prostate segmentation algorithms for mri: the promise12 challenge. Medical
  image analysis  \textbf{18}(2),  359--373 (2014)

\bibitem{feddg}
Liu, Q., Chen, C., Qin, J., Dou, Q., Heng, P.A.: Feddg: Federated domain
  generalization on medical image segmentation via episodic learning in
  continuous frequency space. In: CVPR (2021)

\bibitem{saml}
Liu, Q., Dou, Q., Heng, P.A.: Shape-aware meta-learning for generalizing
  prostate mri segmentation to unseen domains. MICCAI  (2020)

\bibitem{import_phase}
Oppenheim, A., Lim, J.: The importance of phase in signals. Proceedings of the
  IEEE  \textbf{69}(5),  529--541 (1981). \doi{10.1109/PROC.1981.12022}

\bibitem{Orlando2020}
Orlando, J.I., Fu, H., Breda, J.B., van Keer, K., Bathula, D.R., Diaz-Pinto,
  A., Fang, R., Heng, P.A., Kim, J., Lee, J., Lee, J., Li, X., Liu, P., Lu, S.,
  Murugesan, B., Naranjo, V., Phaye, S.S.R., Shankaranarayana, S.M., Sikka, A.,
  Son, J., van~den Hengel, A., Wang, S., Wu, J., Wu, Z., Xu, G., Xu, Y., Yin,
  P., Li, F., Zhang, X., Xu, Y., Bogunovi{\'{c}}, H.: {REFUGE}~challenge: A
  unified framework for evaluating automated~methods for glaucoma~assessment
  from fundus photographs. Medical Image Analysis  \textbf{59},  101570 (Jan
  2020)

\bibitem{unet}
Ronneberger, O., Fischer, P., Brox, T.: U-net: Convolutional networks for
  biomedical image segmentation. In: MICCAI (2015)

\bibitem{sivaswamy2015comprehensive}
Sivaswamy, J., Krishnadas, S., Chakravarty, A., Joshi, G., Tabish, A.S.,
  et~al.: A comprehensive retinal image dataset for the assessment of glaucoma
  from the optic nerve head analysis. JSM Biomedical Imaging Data Papers
  \textbf{2}(1), ~1004 (2015)

\bibitem{mean-teacher}
Tarvainen, A., Valpola, H.: Mean teachers are better role models:
  Weight-averaged consistency targets improve semi-supervised deep learning
  results. In: Advances in Neural Information Processing Systems. vol.~30
  (2017)

\bibitem{vapnik_stat_learn_theory}
Vapnik, V.: Statistical learning theory. Wiley (1998)

\bibitem{dofe}
Wang, S., Yu, L., Li, K., Yang, X., Fu, C.W., Heng, P.A.: {DoFE}:
  Domain-oriented feature embedding for generalizable fundus image segmentation
  on unseen datasets. {IEEE} Transactions on Medical Imaging  \textbf{39}(12),
  4237--4248 (Dec 2020)

\bibitem{fourier_dg}
Xu, Q., Zhang, R., Zhang, Y., Wang, Y., Tian, Q.: A fourier-based framework for
  domain generalization. In: CVPR (2021)

\bibitem{fourier_da}
Yang, Y., Soatto, S.: {FDA:} fourier domain adaptation for semantic
  segmentation. In: CVPR (2020)

\bibitem{cutmix}
Yun, S., Han, D., Chun, S., Oh, S.J., Yoo, Y., Choe, J.: Cutmix: Regularization
  strategy to train strong classifiers with localizable features. In: ICCV
  (2019)

\bibitem{mixup}
Zhang, H., Ciss{\'{e}}, M., Dauphin, Y.N., Lopez{-}Paz, D.: mixup: Beyond
  empirical risk minimization. In: ICLR (2018)

\bibitem{Zhang2020}
Zhang, L., Wang, X., Yang, D., Sanford, T., Harmon, S., Turkbey, B., Wood,
  B.J., Roth, H., Myronenko, A., Xu, D., Xu, Z.: Generalizing deep learning for
  medical image segmentation to unseen domains via deep stacked transformation.
  {IEEE} Transactions on Medical Imaging  \textbf{39}(7),  2531--2540 (Jul
  2020)

\bibitem{zhou2022generalizable}
Zhou, Z., Qi, L., Shi, Y.: Generalizable medical image segmentation via random
  amplitude mixup and domain-specific image restoration. In: ECCV (2022)

\end{thebibliography}
\end{document}